\documentclass[aps,prd,12pt,notitlepage,nofootinbib,tightenlines]{revtex4-1}
\usepackage{amsmath}
\usepackage{amssymb}
\usepackage{epsfig}
\usepackage{color,graphicx}
\usepackage{bm}
\usepackage{times}
\usepackage{braket}
\usepackage{slashed}
\usepackage{hyperref}

\newcommand{\beq}{\begin{eqnarray}}
\newcommand{\eeq}{\end{eqnarray}}

\newcommand{\non}{\nonumber\\ }

\newcommand{\jpsi}{J/\psi}

\newcommand{\psl}{ p \hspace{-2.0truemm}/ }

\newcommand{\epsl}{ \epsilon \hspace{-2.0truemm}/ }

\def \cpc{{ Chin. Phys. C } }
\def \csb{{ Chin. Sci. Bull. } }

\def \epjc{{ Eur. Phys. J. C} }
\def \jhep{ { JHEP } }
\def \jpg{ { J. Phys. G} }

\def \npb{ { Nucl. Phys. B} }
\def \plb{ { Phys. Lett. B} }

\def \prd{ { Phys. Rev. D} }
\def \prl{ { Phys. Rev. Lett.}  }

\def \zpc{ { Z. Phys. C}  }

\definecolor{Blue}{rgb}{0.,0.,1.}

\definecolor{nicegreen}{rgb}{0.1,0.5,0.1}

\hypersetup{colorlinks,citecolor=nicegreen,linkcolor=nicegreen}
\begin{document}

\title{ Semileptonic  decays $B_c \to (\eta_c,\jpsi) l  \bar{\nu}_l $  in the ``PQCD + Lattice"  approach}
\author{Xue-Qing Hu$^{1}$ } \email{hu-xueqing@qq.com}
\author{Su-Ping  Jin$^{1}$ }  \email{2223919088@qq.com}
\author{Zhen-Jun Xiao$^{1,2}$  } \email{xiaozhenjun@njnu.edu.cn}
	\affiliation{1.  Department of Physics and Institute of Theoretical Physics,
		Nanjing Normal University, Nanjing, Jiangsu 210023, People's Republic of China,}
	\affiliation{2. Jiangsu Key Laboratory for Numerical Simulation of Large Scale Complex Systems,
	Nanjing Normal University, Nanjing 210023, People's Republic of China}
	\date{\today}
\begin{abstract}
In this paper, we studied the semileptonic decays $B_c^-  \to (\eta_c, J/\psi) l ^-  \bar{\nu}_l$ by employing the PQCD factorization approach,
using the newly defined distribution amplitudes of the $B_c$ meson and the new kind of parametrization for extrapolation of the form factors ,
and also  taking into account the lattice QCD results about the relevant form factors at several points.
We found the following main results:
(a)   the PQCD  predictions for the branching ratios of  $B_c \to (\eta_c,\jpsi) l \bar{\nu}$ decays  will become a little smaller
 by about $(5-16)\%$  when the lattice input are taken into account in the extrapolation of the relevant form factors;
(b) the PQCD  predictions for   the ratio $R_{\eta_c, J/\psi}$  and the  longitudinal polarization $P_{\tau}$ are
$R_{\eta_c}=0.34\pm 0.01 ,   R_{J/\psi}=0.28\pm 0.01$,
$P_{\tau}(\eta_c) = 0.37\pm 0.01$ and  $P_{\tau}(J/\psi) = -0.55 \pm 0.01$ ;
and  (c)   after the inclusion of the lattice  input the theoretical predictions changed slightly:
$R_{\eta_c}=0.31\pm 0.01$,   $ R_{ J/\psi}=0.27\pm 0.01$,
$P_{\tau}( \eta_c) =  0.36 \pm 0.01$ and  $P_{\tau}( J/\psi) = -0.53\pm 0.01$.
The theoretical predictions for $R_{ J/\psi}$  agree with the measured one within errors,
and  other predictions could be tested in the near future LHCb experiments.
\end{abstract}

\pacs{13.20.He, 12.38.Bx, 14.40.Nd}

\keywords{ The semileptonic $B_c$ decays; the PQCD approach;  Lattice QCD results;  the ratio of the branching ratios; the  longitudinal polarization;  }

\maketitle


\section{ Introduction}\label{sec:1}

In the standard model (SM),  all electroweak gauge bosons  ( $Z, \gamma$ and $W^\pm$)  have equivalent couplings to
three generation leptons,  and the only differences arise due to the mass hierarchy: $m_e <  m_{\mu}  \ll   m_{\tau}$:
this is the so-called Lepton flavor universality (LFU)  in the SM .
The $B_c$ meson can only decay through weak interactions because it is below the B-D threshold, it is therefore an ideal system
to study the weak decays of heavy quarks.
Since the rare semileptonic decays governed by the flavor-changing neutral currents (FCNC) are forbidden
at the tree level in the SM,  the precise measurements for such semileptonic $B_c$ decays can play an important role in testing the SM
and in searching for the signal and/or evidence of the new physics (NP) beyond the SM.
In recent years, the measured values of $R(D)$ and $R(D^*)$ ,  defined as the ratios of the branching fractions
${\cal B}(B \to D^{(*)} \tau \nu_\tau)$ and ${\cal B}(B \to D^{(*)} l \nu_l)$) ,  are clearly larger than the SM predictions:
the combined deviation is   about $3.8\sigma$  for $R(D)-R(D^*)$ in 2017 ~ \cite{hfag2017}, or $3.1\sigma$ after the inclusion
of the newest Belle measurement \cite{Belle1904,Caria1903}.
The semileptonic decays $B \to D^{(*)} l\nu_l$ with $l=(e,\mu,\tau)$  are therefore studied intensively by many authors  in the framework of the SM
~\cite{fajfer-prl109,csb59-125,csb59-3787,prd95-115008,jhep-1711}, or in various new physics (NP) models beyond the SM for example in
Refs.~\cite{prd95-115008,li-16a,adam2019,Fajfer-2018}.

If the above mentioned $R(D^{(*)})$ anomalies are indeed the first signal of the LFU violation ( i.e. an indication of new physics )  in $B_{u,d}$ sector,
it must  appear in the similar semileptonic decays of $B_s$ and $B_c$ mesons,  and  should be studied  systematically.
The $B_c$ ($\bar{b}c$) meson,  as a bound state of  two heavy bottom and charm quarks,   was firstly observed by the CDF Collaboration \cite{CDF1998}
and then by  the Large Hadron Collider (LHC) experiments in recent years \cite{pdg2018}.
The properties of $B_c$ meson and the dynamics involved in $B_c$ decays could be fully exploited through the precise measurements
at the LHC experiments,  especially the measurements carried on by the LHCb Collaboration.
Very recently,  some hadronic  and  semileptonic  $B_c$ meson decays have been measured by LHCb  experiments~\cite{lhcb-17a,lhcb-18a}.
Analogous to the cases for $B$ decays,   the generalization of the $R(D^{(*)})$  for the semileptonic $B_c$ decays  are
 the ratio  $R_{\eta_c}$ and  $R_{\jpsi}$:
\beq
R_X&=& \frac{{\cal B}(B_c^- \to X  \tau^- \bar{\nu}_\tau)}{{\cal B}(B_c^- \to X \mu^-\bar{\nu}_\mu)},  \ \ for \ \ X=(\eta_c,\jpsi).
\label{eq:rdef1}
\eeq
But only the ratio $R_{\jpsi}$ has been measured by the LHCb Collaboration very recently~\cite{lhcb-18a}. ,
\beq
R_{\jpsi}^{\rm Exp}=  0.71 \pm 0.17(stst.) \pm 0.18 (syst.), \label{eq:rdata1}
\eeq
which is consistent with currently available SM predictions
\cite{epjc45-711,cpc37-093102,prd73-094006,epjc51-833,wang09,qiao2013,prd62-014019,jpg26,prd68-094020,prd82-034032,prd48-5208,npb569-473,prd74a}
within $2\sigma$ errors .

During the past two decades, the semileptonic $B_c \to (\eta_c,\jpsi)  l  \bar{\nu}_l$ decays have been studied  by many authors in rather different
theories or models,  for example,   in the QCD sum rule (QCD SR) and  light-cone sum rules (LCSR)~\cite{epjc51-833,prd48-5208,npb569-473, zpc57-43,jhep1905-094},
the relativistic quark model (RQM) or non-relativistic quark model (NRQM) ~\cite{prd68-094020,prd71-094006},
the light-front quark model (LFQM)~\cite{wang09,prd89-017501},
the covariant confining quark model (CCQM)~\cite{prd97-054014},
the nonrelativistic QCD (NRQCD)~\cite{Chang-1992,Qiao-2011,qiao2013,Shen-2014,nrqcd},
the  model-independent investigations (MII)~\cite{mi1,mi2,mi3,z-series},
the lattice QCD (LQCD) ~\cite{lattice1,lattice2,lattice3} and
the perturbative QCD (PQCD) factorization approach \cite{cpc37-093102,epjc76-564,prd96-076001}.

In a previous work ~\cite{cpc37-093102} ,   we  calculated the ratio $R_{\jpsi}$ and $R_{\eta_c}$
 by employing  the PQCD  approach  \cite{pqcd1,pqcd2}   and found the PQCD predictions   ~\cite{cpc37-093102}:
 \beq
 R_{\jpsi} \approx 0.29,  \qquad R_{\eta_c} \approx 0.31, \label{eq:rjpsi0}
 \eeq
which also agree well with the ones  from  the QCDSR or other different approaches in the frame work of the SM .
In this paper, we will present a  new systematic  evaluation for the ratio $R_{\jpsi}$ and $R_{\eta_c}$   by using the
PQCD factorization approach but with the following further improvements:
\begin{enumerate}
\item[(1)]
We here will use a newly developed distribution amplitude (DA) $\phi_{B_c} (x,b)$  for $B_c$ meson as proposed very recently
in Ref.~\cite{Bc-am}:
\beq
\phi_{B_c}(x,b)= \frac{f_{B_c}}{2\sqrt{6}} N_{B_c}x(1-x)\cdot \exp\left[-\frac{(1-x)m_c^2+xm_b^2}{8\beta^2_{B_c}x(1-x)} \right]
\cdot \exp\left[-2\beta^2_{B_c}x(1-x)b^2 \right] , \quad
\label{eq:form2}
\eeq
 instead of the simple $\delta$-function  as being used in Ref.~\cite{epjc45-711,cpc37-093102}:
\beq
 \phi_{B_c} (x)=\frac{f_{B_c}}{2\sqrt{6} } \delta \left (x-\frac{m_c}{m_{B_c}} \right ). \label{eq:form1}
 \eeq

\item[(2)]
For the relevant form factors,  the new preliminary lattice QCD results from the HFQCD Collaboration  include  (a)  the lattice QCD results  for  $V(q^2)$ and $A_1(q^2)$ at
several $q^2$ values for $B_c \to \jpsi$ transition, and  (b)  the lattice QCD results for $f_0(q^2)$ at five $q^2$ values and $f_+(q^2)$ at four
$q^2$ values \cite{lattice1,lattice2}.
We will use four Lattice QCD results  $( f_{0,+}(8.72), V(5.44), A_{1}(10.07)) $  as the  new input in the extrapolation of the relevant  form factors  from the
low $q^2$ region to the $q_{max}^2$.

\item[(3)]
For the extrapolation of the form factors,  analogous to the authors of  Ref.~\cite{jhep1905-094}, we will also use the Bourrely-Caprini-Lellouch (BCL)
parametrization  to make the series expansion of the form factors \cite{bcl09} instead of the exponential expansion formulae as being used in
Ref.~\cite{cpc37-093102}.   We will calculate the branching ratios of the considered decays and the ratios  $R_{\jpsi}$ and $R_{\eta_c}$ by using  the PQCD approach
itself  and the ``PQCD+Lattice'' method respectively,  and compare the resultant predictions.

\item[(4)]
Besides the ratios $R_{\eta_c}$ and $R_{\jpsi}$, we  here will  calculate the longitudinal polarization  $P_\tau(\eta_c)$ and $P_\tau(\jpsi)$
of the final state tau lepton, which was absent in Ref.~\cite{cpc37-093102}.    Just like the polarization $P_\tau^{D^*}$  firstly measured at Belle \cite{prl118-801} ,
both $P_\tau(\eta_c)$ and $P_\tau(\jpsi)$ could  be measured in the future LHCb experiment.

\end{enumerate}

The paper is organized as follows: after this introduction, we give  the distribution amplitudes of the $B_c$ meson
and the final state $\eta_c$ and $\jpsi$ mesons in Section 2.  By employing the PQCD factorization approach we calculate
and present the expressions for the $B_c \to (\eta_c,J/\psi)$ transition form factors in the low $q^2$ region in Section 3.
The extrapolation of the six form factors from the low $q^2$ region to the $q^2_{max}$, the PQCD and the "PQCD+Lattice"  predictions
for the branching ratios ${\cal B}(B_c \to (\eta_c,\jpsi)) (\mu^- \bar{\nu}_\mu, \tau^- \bar{\nu}_{\tau})$,
the ratios $R_{\eta_c}$ and $R_{\jpsi}$ of the branching ratios,  and the longitudinal polarization $P_\tau(\eta_c)$ and $P_\tau(\jpsi)$
are given in Section 4.   A short summary is given in the final section.


\begin{figure}[thb]
\vspace{-2cm}
\centerline{\epsfxsize=15cm \epsffile{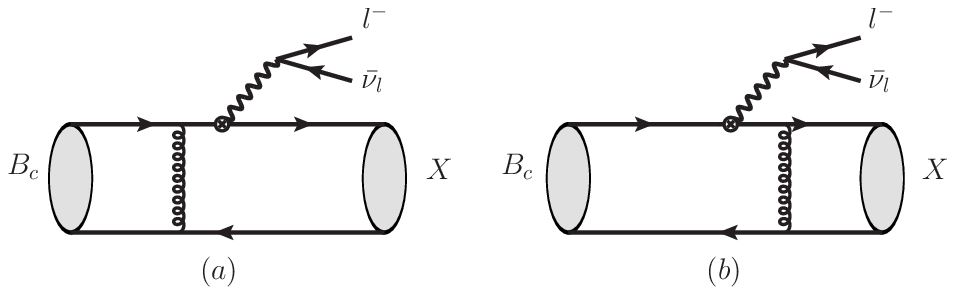}}
\vspace{-16cm}
\caption{ The charged current  tree Feynman diagrams for the semileptonic decays $B_c^- \rightarrow (\eta_c, \jpsi)  l^-\bar{\nu}_l $
with $l= (e,\mu,\tau)$ in the PQCD approach. } \label{fig:fig1}
\end{figure}

\section{ Kinematics and the Wave Functions}\label{sec:2}

The lowest order Feynman diagrams for $B_c \to  X l\nu $ are displayed in Fig.~\ref{fig:fig1}.
The kinematics of these decays are discussed in the large-recoil (low $q^2$) region, where the PQCD factorization approach is applicable to the considered semileptonic
decays involving $\eta_c$ or $\jpsi$ as the final state meson \cite{li1995}.
In the $B_c$ meson rest frame, we define the $B_c$ meson momentum $P_{1}$,
and the final state  meson momentum $P_{2}$ in the light-cone coordinates as\cite{prd67-054028,cpc37-093102}
\beq
\label{eq-mom-p1p2}
P_{\rm 1}=\frac{m_{\rm B_c}}{\sqrt{2}}(1,1,0_\bot),\quad
P_{\rm 2}=r\frac{m_{\rm B_c}}{\sqrt{2}} (\eta^+,\eta^-,0_\bot),
\eeq
with
\beq
\label{eq:eta}
 \eta^{\pm} = \eta \pm \sqrt{\eta^2-1 }, \quad
  \eta =\frac{1}{2r}\left[ 1+ r^2-\frac{q^2}{ m_{B_c}^2 } \right],
\label{eq:eta0}
 \eeq
where the mass ratio $r=m_{\eta_c}/m_{B_c} $ or $m_{\jpsi}/m_{B_c} $,  and $q=p_{\rm 1}-p_{\rm 2}$ is the momentum of the lepton pair.
The longitudinal polarization vector $\epsilon_{\rm L}$ and transverse polarization vector
$\epsilon_{\rm T}$ of the vector meson are defined in the same way as in Ref.~\cite{cpc37-093102}:
\beq
\epsilon_{\rm L}=\frac{1}{\sqrt{2}} (\eta^+,-\eta^-,0_\bot), \qquad \epsilon_{\rm T}=(0,0,1),
\eeq
The momentum $k_1$ and $k_2$ of the spectator quark in $B_c$ or in final state $(\jpsi,\eta_c)$ are parameterized in the same way as
in Ref.~\cite{cpc37-093102}.

For the $B_c$ meson wave function, we make use of the same one as being used for example in Ref.\cite{epjc45-711,cpc37-093102},
\beq
\Phi_{B_c}(x,b)=\frac{i}{\sqrt{6}} (\psl_1 +m_{B_c}) \gamma_5 \phi_{B_c} (x,b).
\label{eq:bc-wa}
\eeq
We here will use  the new DA $\phi_{B_c} (x,b)$ ~\cite{Bc-am} as given in Eq.~(\ref{eq:form2})
instead of the simple $\delta$-function as given in Eq.~(\ref{eq:form1}). As usual, the normalization constant $N_{B_c}$ in Eq.~(\ref{eq:form2})
is fixed by the relation
\beq
\int_0^1 \phi_{B_c}(x,b=0)dx\equiv \int_0^1 \phi_{B_c}(x)dx=\frac{f_{B_c}}{2\sqrt{6}}
\label{eq:Nbc}
\eeq
where the decay constant $f_{B_c}=0.489 \pm 0.005$ GeV has been obtained in lattice QCD by the TWQCD Collaboration~\cite{TW}.
We will set  $\beta_{B_c}=1.0\pm 0.1$ GeV  in Eq.~(\ref{eq:form2})  in order to estimate the uncertainty  ~\cite{Bc-am}.

For the  pseudoscalar  charmonium state $\eta_c$ and the vector one $J/\Psi$,  we use the same wave function as those in
Refs.~\cite{prd73-094006,cpc37-093102}:
\beq
\Phi_{\eta_c}(x) &=&\frac{i}{\sqrt{6}}\gamma_5 \left [ \psl\phi^v(x)+m_{\eta_c}\phi^s(x) \right ], \label{eq:wf-etzc}\\
\Phi_{J/\Psi}^L(x) &=&\frac{1}{\sqrt{6}} \left [ m_{J/\Psi}\epsl_L \phi^L(x)+\epsl_L \psl\phi^{t}(x) \right ]\;,  \label{eq:wf-jpdil}  \\
\Phi_{J/\Psi}^T(x) &=&\frac{1}{\sqrt{6}} \left [ m_{J/\Psi}\epsl_T \phi^V(x)+\epsl_T \psl\phi^{T}(x) \right ]\;, \label{eq:wf-jpsit}
\eeq
where  the  twist-2  asymptotic DAs $ (\phi^v(x),\phi^L(x),\phi^T(x))$ and the twist-3 ones $ (\phi^s(x),\phi^t(x),\phi^V(x))$  are the same
ones as those  being  used in Refs.~\cite{prd73-094006,cpc37-093102}.

\section{ The form factors and differential decay widths}\label{sec:3}

For the considered charged current $B_c\to ( \eta_c,\jpsi ) l^-\bar{\nu}_l$ decays,  the quark-level transition is  the
$b\to c  l^-\bar{\nu}_l$ decay with the effective Hamiltonian
\beq  \label{eq:heff}
{\cal H}_{eff}(b\to c l^-\bar{ \nu}_l)=\frac{G_F}{\sqrt{2}}V_{cb}\;
\bar{c} \gamma_{\mu}(1-\gamma_5)b \cdot \bar l\gamma^{\mu}(1-\gamma_5)\nu_l.
\eeq
where $G_F=1.166 37\times10^{-5} GeV^{-2}$ is the Fermi-coupling constant and $V_{cb}$ is the CKM matrix element.
The differential decay widths of the semi-leptonic decays $B^-_c \to \eta_c  l^-\bar{\nu}_{\rm l}$  can be written   \cite{wang09,cpc37-093102}
in the following form:
\beq
\frac{d\Gamma(B_c \to \eta_c   l\bar{\nu}_{\rm l})}{dq^2}&=&\frac{G_F^2|V_{\rm cb}|^2}{192 \pi^3  m_{\rm B_c}^3}
\left ( 1-\frac{m_{\rm l}^2}{q^2} \right)^2 \frac{\lambda^{1/2}(q^2)}{2q^2} \non
&& \hspace{-1cm} \cdot
 \Bigl \{  3 m_{\rm l}^2\left (m_{\rm B_c}^2-m_{\rm \eta_c}^2 \right )^2 |f_{\rm 0}(q^2)|^2
+ \left (m_{\rm l}^2+2q^2 \right )\lambda(q^2)|f_+(q^2)|^2 \Bigr \}, \label{eq:dg1}
\eeq
where $m_{\rm l}$ is the mass of the charged leptons,  $0\leq q^2 \leq (m_{\rm B_c} -m_{\rm \eta_c})^2 $ and
 $\lambda(q^2) = (m_{\rm B_c}^2+m_{\rm \eta_c }^2-q^2)^2
- 4 m_{\rm B_c}^2 m_{\rm \eta_c }^2 $ is the phase space factor.
In the PQCD factorization approach,  the form factor $f_0(q^2)$ and $f_+(q^2)$ in Eq.~(\ref{eq:dg1})  defined through the matrix element
$<\eta_c(p_2)|\bar{c}(0)\gamma_\mu b(0)|B_c(p_1)>$  \cite{wang09,cpc37-093102}  can be calculated
and written as a summation of the auxiliary form factor $f_{1,2}(q^2)$:
\beq
f_+(q^2)&=&  \frac{1}{2}\left [ f_1(q^2) + f_2(q^2)\right ],  \non
f_0(q^2)&=& F_+(q^2) +  \frac{q^2}{2(m^2_{B_c} - m^2_{\eta_c})}\left [ f_1(q^2)- f_2(q^2) \right ] . \label{eq:f0q2}
\eeq
After making the analytical calculations in PQCD approach, one found  the function $f_{1,2}(q^2)$:
\beq
f_1(q^2)&=&8\pi m_{B_c}^2C_F\int dx_1 dx_2\int b_1 db_1 b_2 db_2
\;\phi_{B_c}(x_1,b_1)\non
&\times &\Bigl\{\left[   -2r^2x_2 \phi^v(x_2)+2r(2-r_b)\phi^s(x_2)   \right]  \cdot  H_1(t_1) \non
&+& \left[ \left ( -2r^2+\frac{rx_1\eta^+\eta^+}{\sqrt{\eta^2-1}} \right )\phi^v(x_2)
+\left ( 4rr_c-\frac{2x_1r\eta^+}{\sqrt{\eta^2-1}} \right )\phi^s(x_2)\right]  \cdot H_2(t_2)  \Bigr \},
\label{eq:f1q2}
\eeq
\beq
f_2(q^2)&=&8\pi m_{B_c}^2C_F\int dx_1 dx_2\int b_1 db_1 b_2 db_2 \;\phi_{B_c}(x_1,b_1)\non
&\times& \Bigl\{ \left[(4r_b-2+4x_2r\eta)\phi^v(x_2)+(-4rx_2)\phi^s(x_2)\right] \cdot  H_1(t_1) \non
&+& \left[ \left (-2r_c-\frac{x_1\eta^+}{\sqrt{\eta^2-1}} \right )\phi^v(x_2)
+\left ( 4r+\frac{2x_1}{\sqrt{\eta^2-1}} \right )\phi^s(x_2)\right]  \cdot  H_2(t_2) \Bigr \}, \label{eq:f2q2}
\eeq
with the functions $H_i(t_i)$ can be written in the following form
\beq
H_i(t_i)=h_i(x_1,x_2,b_1,b_2) \cdot \alpha_s(t_i) \exp\left [-S_{ab}(t_i) \right ], \quad {\rm  for} \quad   i=(1,2) ,  \label{eq:hiti}
\eeq
where $C_F=4/3$ is a color factor, $r_c=m_c/m_{B_c}$, $r_b=m_b/m_{B_c}$, $r=m_{\eta_c}/m_{B_c}$ .
The explicit expressions of  the hard functions $h_i(x_1,x_2,b_1,b_2)$ and the Sudakov functions $\exp\left [-S_{ab}(t_i) \right ] $ will be given in Appendix.

For $B^-_c \to \jpsi  l^-\bar{\nu}_{\rm l}$ decays,  the differential decay widths can be written   in the following form  \cite{wang09,cpc37-093102}:
\beq
\frac{d\Gamma_{\rm L}}{dq^{\rm 2}}&=&
\frac{G_{\rm F}^{\rm 2}|V_{\rm cb}|^{\rm 2}}{192 \pi^3  m_{\rm B_c}^3} \left ( 1-\frac{m_{\rm l}^{\rm 2}}{q^{\rm 2}}\right )^{\rm 2}
\frac{\lambda^{1/2}(q^{\rm 2})}{2q^{\rm 2}}\cdot \Bigg\{3m^{\rm 2}_{\rm l}\lambda(q^{\rm 2})A^{\rm 2}_{\rm 0}(q^{\rm 2})\non
&& \hspace{-1cm}
+\frac{m^{\rm 2}_{\rm l}+2q^{\rm 2}}{4m^{\rm 2}_{\jpsi }}\cdot \left [(m^{\rm 2}_{\rm B_c}-m^{\rm 2}_{\jpsi }-q^{\rm 2})(m_{\rm B_c}+m_{\jpsi})
A_{\rm 1}(q^{\rm 2}) -\frac{\lambda(q^{\rm 2})}{m_{\rm B_c}+m_{\jpsi }}A_{\rm 2}(q^{\rm 2}) \right ]^{\rm 2} \Bigg\},
\label{eq:dfds1}
\eeq
\beq
\frac{d\Gamma_\pm}{dq^{\rm 2}}&=&
\frac{G_F^{\rm 2}|V_{\rm cb}|^{\rm 2}}{192 \pi^3
 m_{\rm B_c}^3}\left ( 1-\frac{m_{\rm l}^{\rm 2}}{q^{\rm 2}}\right )^{\rm 2} \frac{\lambda^{3/2}(q^{\rm 2})}{2}\non
& & \cdot \left \{ (m^{\rm 2}_{\rm l}+2q^{\rm 2})\left[\frac{V(q^{\rm 2})}{m_{\rm B_c}+m_{\jpsi }}\mp
\frac{(m_{\rm B_c}+m_{\jpsi })A_{\rm 1}(q^{\rm 2})}{\sqrt{\lambda(q^{\rm 2})}}\right]^{\rm 2}\right\},
\label{eq:dfds2}
\eeq
where  $0\leq q^2 \leq (m_{\rm B_c} -m_{\rm \jpsi})^2 $ and
$\lambda(q^2) = (m_{\rm B_c}^2+m_{\jpsi }^2-q^2)^2 - 4 m_{\rm B_c}^2 m_{\jpsi}^2$ .
The total differential decay widths is defined as
\beq
\frac{d\Gamma}{dq^{\rm 2}}=\frac{d\Gamma_{\rm L}}{dq^{\rm 2}} + \frac{d\Gamma_+}{dq^{\rm 2}}    +\frac{d\Gamma_-}{dq^{\rm 2}}\; .
\label{eq:dfdst}
\eeq
The form factors $V(q^2)$ and $A_{0,1,2}(q^2)$  can also be calculated in the framework
of the PQCD factorization approach:
{\small
\beq  
V(q^2)&=&8\pi m_{B_c}^2C_F\int dx_1 dx_2\int b_1 db_1 b_2 db_2
 \;\phi_{B_c}(x_1,b_1)\cdot (1+r)\non
&\times & \left  \{\left[(2-r_b)\phi^T(x_2)-rx_2\phi^V(x_2)\right] \cdot H_1(t_1)
+ \left[\left(r+\frac{x_1}{2\sqrt{\eta^2-1}}\right)\phi^V(x_2) \right]\cdot H_2(t_2)  \right  \},
\label{eq:Vqq}
\eeq
\beq
A_0(q^2)&=&8\pi m_{B_c}^2C_F\int dx_1 dx_2\int b_1 db_1 b_2 db_2
 \;\phi_{B_c}(x_1,b_1)\non
&\times & \Bigl \{  \left[\left(2r_b-1-r^2x_2+2rx_2\eta\right)\phi^L(x_2)
+r\left(2-r_b-2x_2\right)\phi^t(x_2) \right] \cdot H_1(t_1)  \non
&+& \left [\left(r^2+r_c+\frac{x_1}{2}-rx_1\eta
+\frac{x_1(\eta+r(1-2\eta^2))}{2\sqrt{\eta^2-1}}
 \right)\phi^L(x_2)\right ] \cdot H_2(t_2)  \Bigr \},
\label{eq:A0qq}
\eeq
\beq
A_1(q^2)&=&8\pi m_{B_c}^2C_F\int dx_1 dx_2\int b_1 db_1 b_2 db_2
\;\phi_{B_c}(x_1,b_1)\cdot \frac{r}{1+r}\non
&\times & \Bigl \{\left [ 2(2r_b-1+rx_2\eta)\phi^V(x_2)
-2(2rx_2-(2-r_b)\eta)\phi^T(x_2) \right ] \cdot H_1(t_1) \non
& + &\left[ \left(2r_c-x_1+2r\eta \right) \phi^V(x_2)\right]\cdot H_2(t_2) \Bigr \},
\label{eq:A1qq}
\eeq
\beq
A_2(q^2)&=&\frac{(1+r)^2(\eta-r)}{2r(\eta^2-1)}\cdot A_1(q^2)-
8\pi m_{B_c}^2C_F\int dx_1 dx_2\int b_1 db_1 b_2 db_2
\;\phi_{B_c}(x_1,b_1)\cdot \frac{1+r}{\eta^2-1}\non
&\times & \Bigl \{  \left[ \left [  2x_2r(r-\eta)+(2-r_b)(1-r\eta) \right ]\phi^t(x_2) \right. \non
&&\left.  + \left [ (1-2r_b)(r-\eta)-rx_2+2x_2r\eta^2-x_2r^2\eta \right ] \phi^L(x_2)\right] \cdot H_1(t_1)  \non
&+& \left[ x_1\left(r\eta-\frac12\right)\sqrt{\eta^2-1}
+\left(r_c-r^2-\frac{x_1}{2}\right)\eta
+r\left(1-r_c-\frac{x_1}{2}+x_1\eta^2\right) \right] \non
&& \cdot \phi^L(x_2) \cdot H_2(t_2)  \Bigr \},
\label{eq:A2qq}
\eeq }
where $r_c=m_c/m_{B_c}$, $r_b=m_b/m_{B_c}$ and $r=m_{\jpsi}/m_{B_c}$,  the parameter $\eta$ is defined in Eq.~(\ref{eq:eta0}),
and the functions  $H_i(t_i)$ are the same  ones as those defined  in Eq.~(\ref{eq:hiti}).

\section*{4 \hspace{0.3cm} Numerical Results} \label{sec:4}

In the numerical calculations we use the following input parameters
(here masses and decay constants are in units of GeV)\cite{hfag2017,pdg2018,TW}:
\beq
m_{\rm B_c}&=&6.275,\quad  m_{\rm \jpsi}=3.097,\quad m_{\tau}=1.777,  \quad m_{c}=1.27 \pm 0.03,\quad m_{\rm \eta_c}=2.983,\non
 \tau_{\rm B_c}&=&0.507\; {\rm ps},\quad f_{\rm B_c}=0.489\pm 0.005, \quad f_{\rm \eta_c}=0.438\pm 0.008, \quad f_{\rm \jpsi}=0.405 \pm 0.014, \non
|V_{\rm cb}|&=&(42.2 \pm 0.8)\times 10^{-3},  \quad \Lambda^{\rm (f=4)}_{\overline{\rm MS}} = 0.287.
\label{eq:inputs}
\eeq

\begin{table}[thb]
\begin{center}
\caption{The theoretical  predictions for the form factors $f_{\rm 0,+}, V$ and
$A_{\rm 0,1,2}$ at $q^{\rm 2}=0$,  obtained by employing the PQCD approach,  by using some other different approaches
\cite{wang09,BSW,nrqcd,epjc51-833,jhep1905-094,prd68-094020,prd97-054014} or in Lattice QCD \cite{lattice1} .} \label{tab:ff1}
\vspace{0.2cm}
\begin{tabular}{l| c c |c c c c cccc} \hline
Form factors&PQCD & PQCD &LFQM                & BSW              & NRQCD           & LCSR & LCSR & RQM&CCQM& Lattice\\
\ \ \ &This work &\cite{cpc37-093102} & \cite{wang09} & \cite{BSW} & \cite{nrqcd} & \cite{epjc51-833}& \cite{jhep1905-094}&\cite{prd68-094020}
& \cite{prd97-054014} &\cite{lattice1}   \\ \hline
$f_{\rm 0,+} ^{\rm B_c\to \eta_c}(0)$& $ 0.56(7) $ &$0.48(7)$&$ 0.61 $& $ 0.58 $& $ 1.67 $& $ 0.87 $ &  $0.62$& $0.47$ & $0.75$  & $0.59$ \\ \hline
$V ^{\rm B_c\to \jpsi}(0)$                       & $ 0.75(9) $ &$0.42(2)$&$ 0.74 $& $ 0.91 $& $ 2.24 $& $ 1.69 $ & $0.73$ & $0.49$ & $0.78$  & $0.70$  \\
$A_{\rm 0} ^{\rm B_c\to \jpsi}(0)$      & $ 0.40(5) $ &$0.59(3)$&$ 0.53 $&$ 0.58 $& $ 1.43 $& $ 0.27 $ & $0.54$ &  $0.40$ & $0.56$  & $-$  \\
$A_{\rm 1}^{\rm B_c\to \jpsi} (0)$      & $ 0.47(5) $ &$0.46(3)$&$ 0.50 $&$ 0.63 $& $ 1.57 $& $ 0.75 $ & $0.55$ &  $0.73$ & $0.55$  & $0.48$ \\
$A_{\rm 2}^{\rm B_c\to \jpsi} (0)$      & $ 0.62(6) $ &$0.64(3)$&$ 0.44 $&$ 0.74 $& $ 1.73 $& $ 1.69 $ & $0.35$ &  $0.50$ & $0.56$  & $-$ \\
\hline \hline
\end{tabular}
\end{center} \end{table}

For the considered  semileptonic $B_c$ meson decays,  it is easy to see  that the theoretical  predictions for the differential decay rates  and  other physical
observables  strongly depend on the form factors $f_{\rm 0,+}(q^{\rm 2})$  for $B_c \to \eta_c  l \nu_l$ decays ,
and  the form factors $V(q^{\rm 2})$ and $A_{\rm 0,1,2}(q^{\rm 2})$ for $B_c \to \jpsi  l \nu_l$ decays \cite{wang09,cpc37-093102}.
The value of these form factors at $q^2=0$ and their $q^2$ dependence in the whole range of $0\leq  q^2 \leq q^2_{max}$ contain
a lot of information of the physical process.
Up to now, these form factors have been calculated in many rather different methods, for example,  in
Refs.~\cite{zpc57-43,prd48-5208,npb569-473, epjc51-833,jpg26,prd68-094020,prd71-094006}.

In Refs.~\cite{csb59-125,csb59-3787,prd67-054028,prd78-014018},  the authors examined the applicability of the PQCD factorization approach to
$(B \to D^{(*)})$ transitions, and have shown that the PQCD approach with the inclusion of the Sudakov effects is  applicable to study the
semileptonic decays $B \to D^{(*)} l\bar{\nu}_{\rm l}$  \cite{csb59-125,csb59-3787}.
Since  the PQCD predictions for the considered form factors are reliable only at the low $q^{\rm 2}$ region, we  first calculate explicitly
the values of the relevant form factors   at the sixteen points  in the lower region  $ 0 \leq q^{\rm 2} \leq m_{\rm \tau}^{\rm 2}$
by  using the expressions as given in Eqs.~(\ref{eq:f1q2},\ref{eq:f2q2},\ref{eq:Vqq}-\ref{eq:A2qq}) and the definitions in Eq.~(\ref{eq:f0q2}).
 In the second column of  Table \ref{tab:ff1}, we show the PQCD predictions for six relevant form factors at $q^2=0$.
The errors of the PQCD predictions are the combination of the major errors from the uncertainty of $\beta_{\rm B_c}=1.0\pm 0.1$ GeV,
$m_{\rm c}=1.27 \pm 0.03$ GeV and  $|V_{cb}|= (42.2 \pm 0.8)\times 10^{-3}$.
In the third column of  Table \ref{tab:ff1}, we  show the previous PQCD predictions presented in Ref.~\cite{cpc37-093102}.
As a comparison, we also list the central values of the  form factors $f_i(0)$ obtained in some other different  approaches:
such as the BSW \cite{BSW},  the NRQCD \cite{nrqcd},  the LCSR \cite{epjc51-833,jhep1905-094},  RQM and CCQM msthods \cite{prd68-094020,prd97-054014}
or from the lattice QCD \cite{lattice1}.

It is easy to see from the numerical values as given in Table \ref{tab:ff1} that (a) the PQCD predictions for $f_{0,+}(0)$, $V(0)$ and $A_1(0)$ agree very well with the
corresponding lattice QCD results, and (b) the theoretical predictions from different approaches can also be rather different in values, by a factor of three for $f_{0,+}(0)$ for instance.
Since the PQCD calculations for form factor are not reliable in the large  $q^{\rm 2}$ region ,  we have to make an extrapolation for all relevant
form factors from the lower $q^{\rm 2}$ region to the larger $q^{\rm 2}$ region.
In this work we will make the extrapolation by using  the following two different methods.

In the first method,  we use our PQCD predictions for all relevant form factors $f_i(q^2)$ at the sixteen points of $0\leq q^2 \leq m^2_\tau$  as input,
and then make the extrapolation from  low $q^2$ region  to the $q_{\rm max}^2$ by using  the Bourrely-Caprini-Lellouch (BCL) parametrization
\cite{bcl09}.  Analogous to Ref.~\cite{jhep1905-094}, we here also consider only the first two terms  of the series  in the parameter $z$:
\begin{eqnarray}
f_i(t) &=& \frac{1}{1-t/m^2_R} \sum_{k=0}^{1} \alpha^i_k\; z^k(t,t_0)\non
&=&\frac{1}{1-t/m^2_R} \left ( \alpha^i_0 + \alpha^i_1\;    \frac{ \sqrt{t_+ - t} - \sqrt{t_+ - t_0}}{\sqrt{t_+ - t} + \sqrt{t_+ - t_0}} \right ) ,
\label{eq:extra1}
\end{eqnarray}
where $t=q^2$, $m_R$ are the masses of the low-laying $B_c$ resonance as listed in Table \ref{tab:ff2}, and the parameters  $t_\pm$ and $t_0$
\footnote{We here use the same optimized value of $t_0$ as the one
in Refs.~\cite{bfw10,jhep1905-094} without further discussion. }   are as follows:
\beq
0 \leq t_{0} &=&  t_+  \left (1 - \sqrt{1 - \frac{t_-}{t_+}} \right ) \leq t_-, \non
t_{\pm} &=& (m_{B_c} \pm m_x)^2 ,  \label{eq:t0}
\eeq
where $m_x=m_{\eta_c}$ or $m_{\jpsi}$ for $B_c \to \eta_c$ and $\jpsi$ transition, respectively.
In Table  \ref{tab:ff2},  we list the PQCD input $f_i(0)$,  the masses $m_R$,   the parameters
$\alpha_0$ and $\alpha_1$ determined from the BCL fitting procedure for $B_c \to \eta_c$  and $B_c \to J/\psi$ form factors.
The values of  $m_R$ are taken  from Ref.~\cite{jhep1905-094} directly.

\begin{table}[thb]
\begin{center}
\caption{The  form factors $f_i(0)$ from the PQCD calculations, the $J^P$ and masses (in unit of GeV) of the low-laying $B_c$ resonances \cite{jhep1905-094}
used in the BCL fit for $B_c \to (\eta_c,\jpsi)$ form factors.  The parameters $\alpha_{0,1}$ are determined from the fitting. }
\label{tab:ff2}
\vspace{0.2cm}
 \begin{tabular}{|c | c|c c | l l |}
 \hline
FFs &  $f_i(0)$ in PQCD  & $J^P$ & $m_R$ & $\alpha_0$ & $\alpha_1$ \\
\hline
$f_0$ &0.56(7) &$0^+$ & 6.71 & 0.691 & -7.74 \\
$f_+$ & 0.56(7)&$1^-$ & 6.34 & 0.763 & -12.2 \\
\hline
$V$   & 0.75(9)& $1^-$ & 6.34 & 1.06 & -20.6 \\
$A_0$ & 0.40(5)& $0^-$ & 6.28 & 0.551 & -10.5 \\
$A_1$ & 0.47(5)& $1^+$ & 6.75 & 0.586 & -7.73 \\
$A_2$ & 0.62(6)& $1^+$ & 6.75 & 1.01 & -26.8 \\
\hline
\end{tabular}\end{center}
\end{table}

The second method is the ``PQCD+Lattice" method,  similar with what we did in Ref.~\cite{sb60-2009} for the studies of $R(D^{*})$.
As mentioned in the introduction, the authors in HPQCD Collaboration \cite{lattice1,lattice2}  calculated the form factors $f_{0,+}(q^2)$ for $B_c \to \eta_c$ transition,
and $V(q^2)$ and $A_1(q^2)$   for  $B_c \to \jpsi$ transition by using the  lattice  QCD method ( working directly at $m_b$ with an improved non-relativiatic QCD (NRQCD) effective theory formulism )  at  $q^2=0$ and several  other points of $q^2$.
In order to improve the reliability of the extrapolation  of $f_i(q^2)$ to the larger $q^2$ region,  we use   currently available ``Lattice'' results  at points
$q^2=(5.44, 8.72,10.07)$ GeV$^2$ as given in Refs.~\cite{lattice1,lattice2},
\beq
f_0(8.72) &=& 0.823\pm 0.050, \quad  f_+(8.72) = 0.995\pm 0.050,   \non
V(5.44) &=& 1.06\pm 0.05, \quad A_1(10.07) = 0.788\pm 0.050, \label{eq:lattices}
\eeq
as the lattice QCD input in the fitting process for the form factors $(f_{0,+}(q^2), V(q^2), A_1(q^2))$. In order to estimate the effects of possible
uncertainties of the lattice QCD inputs, we here set a roughly five percent error ( $\pm 0.05$ ) to the four form factors in Eq.~(\ref{eq:lattices}).
For other two  form factors $A_{0}(q^2)$ and $A_{2}(q^2)$, unfortunately,   no lattice QCD results are available  at present.

\begin{figure}[thb]
\centerline{\epsfxsize=7.3cm\epsffile{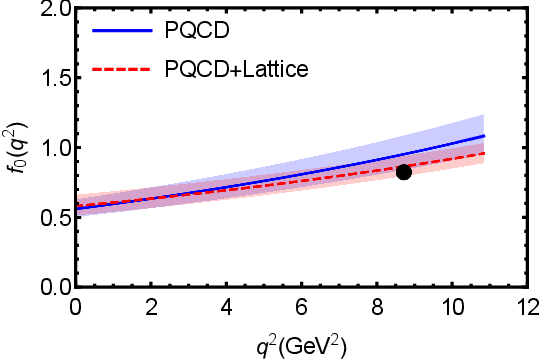}\hspace{0.5cm} \epsfxsize=7.3cm \epsffile{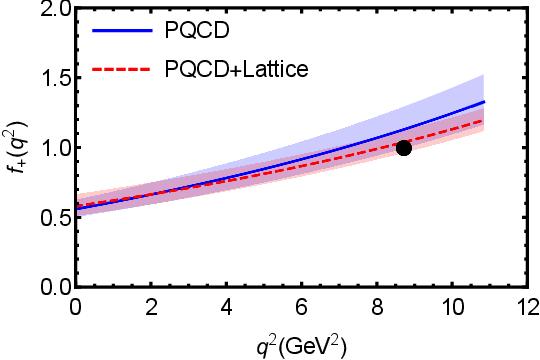} }
\caption{ (Color online) The theoretical predictions for  $B_c \to \eta_c$  transition form factors $f_+(q^2)$ and $f_0(q^2)$ in the PQCD approach ( the blue solid
curve), and the ``PQCD+Lattice" approach ( the red dashed curve). The large dot symbols are the lattice QCD inputs as listed in Eq.~(\ref{eq:lattices})  \label{fig:fig2} }
\end{figure}

\begin{figure}[]
\centerline{\epsfxsize=7.3cm\epsffile{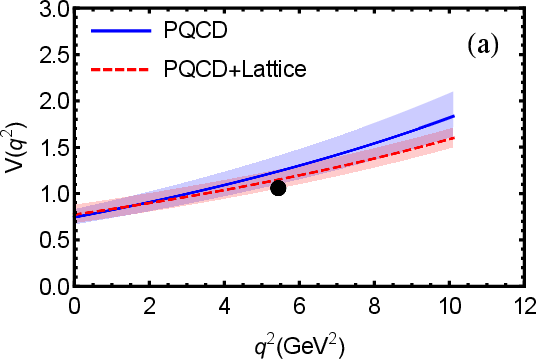}\hspace{0.5cm} \epsfxsize=7.3cm \epsffile{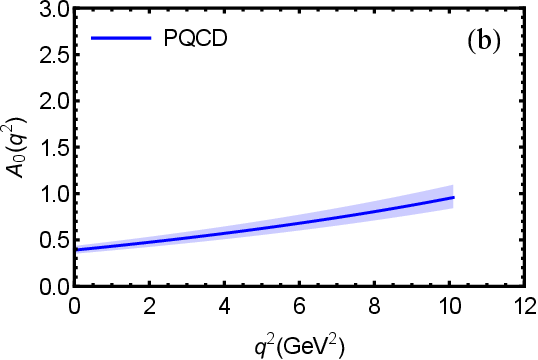}  }
\vspace{0.5cm}
\centerline{ \epsfxsize=7.3cm\epsffile{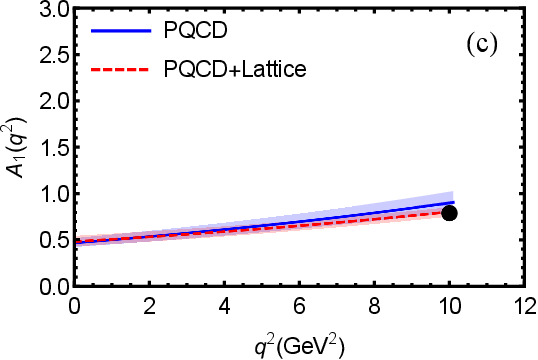}\hspace{0.5cm} \epsfxsize=7.3cm \epsffile{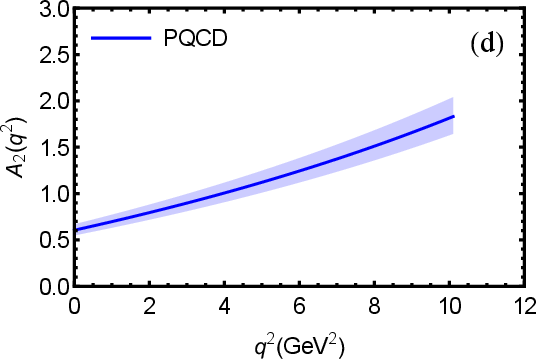}  }
\caption{ (Color online) The theoretical predictions for   $B_c \to \jpsi$ transition form factors $V(q^2)$ and $A_{0,1,2}(q^2)$ in the
 PQCD approach ( the blue solid curve), and the ``PQCD+Lattice" approach ( the red dashed curve).
 The large dot symbols in (a,c) are the lattice QCD inputs as  listed in Eq.~(\ref{eq:lattices}). \label{fig:fig3} }
\end{figure}

In Figs.~\ref{fig:fig2} and \ref{fig:fig3} ,  we show  the theoretical predictions for the  $q^2$-dependence of the six relevant form factors for $B_c \to (\eta_c,\jpsi)$
transitions,  obtained by using the PQCD approach and   the ``PQCD+ Lattice" approach, respectively.
In these two figures,  the blue solid  curves  show the theoretical  predictions for the $q^2$-dependence of   $f_{0,+}(q^2)$, $V(q^2)$ and $A_{0,1,2}(q^2)$
in the PQCD approach,   while  the red dashed curves show the  four form factors $(f_{0,+}(q^2), V(q^2), A_1(q^2))$  obtained  by
using the "PQCD+Lattice" approach. The band in these figures show the uncertainties of the corresponding theoretical predictions for the form factors.
The four black dots  symbols in Fig.~\ref{fig:fig2} and \ref{fig:fig3} show the lattice QCD input in Eq.~\ref{eq:lattices} used in the fitting procedure.
One can see from  the theoretical predictions as illustrated in Figs.~\ref{fig:fig2} and \ref{fig:fig3} that
the values and their $q^2$-dependence of  the form factors obtained by using the two different methods agree very well with each other
in  the whole range of $q^2$.

\begin{figure}[thb]
\centerline{\epsfxsize=7.3cm\epsffile{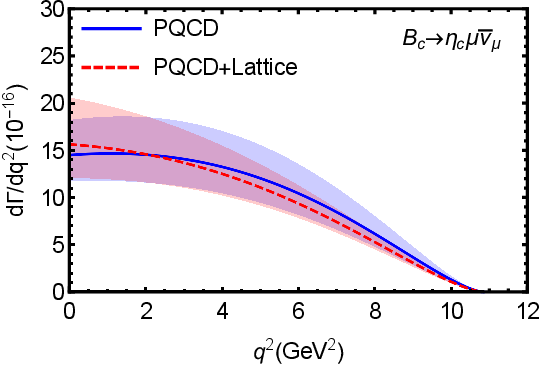}\hspace{0.5cm} \epsfxsize=7.3cm \epsffile{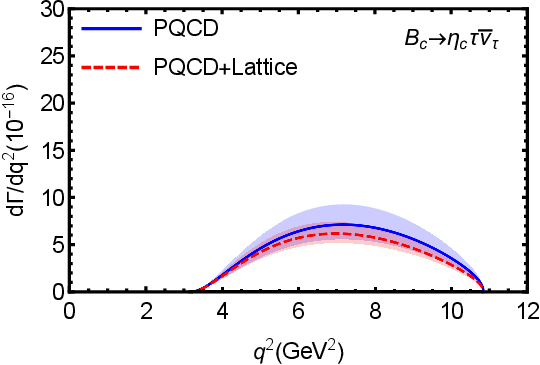}  }
\vspace{1cm}
\centerline{ \epsfxsize=7.3cm\epsffile{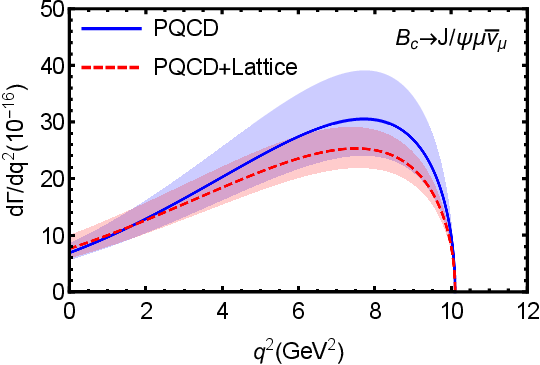}\hspace{0.5cm} \epsfxsize=7.3cm \epsffile{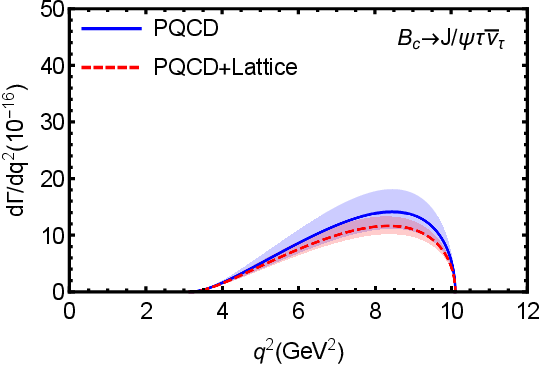}  }
\caption{ (Color online) The theoretical predictions for the $q^2$ dependence of $d\Gamma /dq^2$ for the considered  decays
$B_c \to (\eta_c,\jpsi) (\mu  \bar{\nu}_{\mu}, \tau  \bar{\nu}_{\tau}) $ in both the PQCD  and the  ``PQCD+Lattice'' approaches.
The bands show the theoretical uncertainties. }
\label{fig:fig4}
\end{figure}

In Fig.~\ref{fig:fig4},   we show the $q^2$-dependence of the theoretical predictions for the differential decay rates $d\Gamma /dq^2$ for the semileptonic decays
$B_c \to (\eta_c,\jpsi) l  \bar{\nu}_l $ with $l=(\mu,\tau)$,  where the blue solid  curve and the red dashed ones   show the $d\Gamma /dq^2$ in the
PQCD approach and ``PQCD+Lattice" method respectively.
For the four considered $B_c \to (\eta_c,\jpsi)( \mu^-  \bar{\nu}_{\mu},\tau^-\bar{\nu}_{\tau} ) $ decays,  the theoretical
predictions for the differential decay rates from the PQCD and the ``PQCD+Lattice" approach agree well within errors  in the whole $q^2$ region.
For $B_c \to \jpsi  \mu^-  \bar{\nu}_{\mu} $ decay,  on the other hand, the difference between the central values is a little evident in the large $q^2$
region but remains small in size.
We do wish the  lattice  results for   the form factors $A_{0,2}(q^2)$ become available soon and which will help us to improve our work.

From the formulae of the differential decay rates as given in Eqs.~(\ref{eq:dg1},\ref{eq:dfdst}),
it is straightforward to make the integration over the range of $ m^2_l \le q^2 \le (m^2_{B_c} - m_x^2)$ with
$x = (\eta_c,\jpsi) $. The theoretical predictions (in unit of $10^{-3}$) for the branching ratios of the considered semileptonic decays  are the following:
\beq
{\cal B}(B_c \to \eta_c \tau \bar{\nu}_\tau) &=& \left \{ \begin{array}{ll}
 2.79^{+0.83}_{-0.61}(\beta_{\rm B_c}) \pm 0.11(V_{\rm cb}) \pm 0.09 (m_{\rm c}) ,  &  {\rm PQCD}, \\
 2.41^{+0.48}_{-0.39}(\beta_{\rm B_c})\pm 0.09(V_{\rm cb}) \pm 0.04 (m_{\rm c}) ,   & {\rm PQCD+Lattice}, \\ \end{array} \right.   \label{eq:br11}  \\
 {\cal B}(B_c \to \eta_c \mu \bar{\nu}_\mu ) &=& \left \{ \begin{array}{ll}
 8.14^{+1.91}_{-1.72}(\beta_{\rm B_c}) \pm 0.31(V_{\rm cb}) \pm 0.30 (m_{\rm c}) ,   &  {\rm PQCD}, \\
 7.76^{+1.92}_{-1.46}(\beta_{\rm B_c})\pm 0.29(V_{\rm cb}) \pm 0.24 (m_{\rm c})  ,   & {\rm PQCD+Lattice}, \\ \end{array} \right.  \label{eq:br12}
\eeq
\beq
 {\cal B}(B_c \to \jpsi \tau \bar{\nu}_{\tau}) &=&  \left \{ \begin{array}{ll}
 4.54^{+1.27}_{-0.98}(\beta_{\rm B_c})\pm 0.18(V_{\rm cb}) \pm 0.16 (m_{\rm c}) ,  &  {\rm PQCD}, \\
 3.83^{+0.61}_{-0.55}(\beta_{\rm B_c})\pm 0.14(V_{\rm cb}) \pm 0.10 (m_{\rm c})  ,   & {\rm PQCD+Lattice}, \\ \end{array} \right.  \label{eq:br21}  \\
 {\cal B}(B_c \to \jpsi  \mu \bar{\nu}_{\mu}) &=&   \left \{ \begin{array}{ll}
 16.1^{+4.4}_{-3.3}(\beta_{\rm B_c})\pm 0.61 (V_{\rm cb}) \pm 0.52 (m_{\rm c})   ,  &  {\rm PQCD}, \\
 14.1^{+2.6}_{-2.1}(\beta_{\rm B_c})\pm 0.51(V_{\rm cb}) \pm 0.36 (m_{\rm c})   ,   & {\rm PQCD+Lattice}, \\ \end{array} \right.   \label{eq:br22}
 \eeq
where the major theoretical errors come from the uncertainties of the input parameters $\beta_{\rm B_c}=1.0\pm 0.1$ GeV,
$|V_{\rm cb}|=(42.2\pm 0.8) \times 10^{-3}$ and $m_{\rm c}=1.27\pm 0.03$ GeV.


\begin{table}[]
\begin{center}
\caption{The theoretical  predictions (in unit of $10^{-3}$) for the branching ratios  ${\cal B}(B_c \to (\eta_c,\jpsi ) l \bar{\nu}_l)$
in the PQCD and ``PQCD+Lattice" approaches.  As a comparison, we also list the predictions  as given in a previous PQCD work
\cite{cpc37-093102}, and other four approaches \cite{wang09,epjc51-833,z-series,jhep1905-094}.}
\label{tab:br1a}
\vspace{0.2cm}
\begin{tabular}{l| ll| c c c c c} \hline \hline
Mode  &                                                                                    PQCD &  PQCD+Lattice & PQCD                                & LFQM                   & Z-Series             & LCSR                          &LCSR \\
             &                                                                                                &                              &  \cite{cpc37-093102} & \cite{wang09}  & \cite{z-series} & \cite{epjc51-833} &\cite{jhep1905-094} \\ \hline
${\cal B}(B_c  \to \eta_c\mu \bar{\nu}_\mu)$ & $ 8.14^{+1.96}_{-1.77} $&$7.76^{+1.95}_{-1.51}$         &$ 4.4^{+1.2}_{-1.1} $              &$ 6.7 $                           & $6.6$    & $16.7$ & $8.2^{+1.2}_{-1.1}$\\
${\cal B}(B_c  \to \eta_c\tau \bar{\nu}_\tau)$ & $ 2.79^{+0.84}_{-0.63} $&$2.41^{+0.49}_{-0.40}$         &$ 1.4^{+0.4}_{-0.3} $              &$ 1.9 $                           & $2.0$    & $4.9$   & $2.6^{+0.6}_{-0.5}$ \\ \hline
${\cal B}(B_c  \to \jpsi \mu  \bar{\nu}_\mu) $ & $ 16.1^{+4.5}_{-3.4} $      &$14.1^{+2.7}_{-2.2}$         &$ 10.0^{+1.3}_{-1.2} $             &$ 14.9 $                         & $14.5$ & $23.7$&$22.4^{+5.7}_{-4.9}$ \\
${\cal B}(B_c  \to \jpsi\tau \bar{\nu}_\tau )  $ & $ 4.54^{+1.29}_{-1.01} $  &$3.83^{+0.63}_{-0.58}$         &$ 2.9 ^{+0.4}_{-0.3}$              &$ 3.7 $                           & $3.   6$ & $6.5$ & $5.3^{+1.6}_{-1.4}$ \\
\hline \hline
\end{tabular} \end{center} \end{table}

\begin{table}[]
\begin{center}
\caption{The theoretical  predictions for the ratios $R_{\eta_c}$ and $R_{\jpsi}$ obtained by employing  the PQCD and ``PQCD+Lattice" approaches,
 or as given in  previous works  \cite{cpc37-093102,wang09,epjc51-833,z-series,jhep1905-094,mi1,mi2}. }
\label{tab:br1b}
\vspace{0.2cm}
\begin{tabular}{l| l l| c c c c cc} \hline \hline
Mode  &                                                                                    PQCD &  PQCD+Lattice & PQCD   & LFQM  & Z-Series   & LCSR  &LCSR & M-Ind.\\
             &                                                  &                              &  \cite{cpc37-093102} & \cite{wang09}  & \cite{z-series} & \cite{epjc51-833}
             &\cite{jhep1905-094}& \cite{mi1,mi2} \\ \hline
$R_{\eta_c}$ & $ 0.34\pm 0.01 $ &$0.31\pm 0.01$  &$0.31$        &$ 0.28 $ & $0.31$ & $0.30$&$0.32\pm 0.02$ & $0.29\pm 0.05$ \\
$R_{\jpsi}   $ & $ 0.28\pm 0.01 $ &$0.27\pm 0.01$  &$0.29$        &$ 0.25 $ & $0.25$ & $0.27$& $0.23\pm 0.01$ & $[0.20, 0.39]$ \\
\hline\hline
\end{tabular}
\end{center} \end{table}

In Table \ref{tab:br1a},  we list the theoretical  predictions (in unit of $10^{-3}$) for the branching ratios  of the
considered decays $B_c \to (\eta_c,\jpsi ) l^- \bar{\nu}_l$ with $l=(\mu,\tau)$,  obtained in this paper by employing the
PQCD and the ``PQCD+Lattice" approaches. And as a comparison, we also show the results from the previous PQCD work
\cite{cpc37-093102}, and from several  different models or approaches \cite{wang09,epjc51-833,z-series,jhep1905-094}.
One can see that the difference between different theoretical predictions  can be as large as a factor of two for the same decay mode.
In Table \ref{tab:br1b},  we show the theoretical  predictions  for the ratios  $R_{\eta_c}$ and $R_{\jpsi}$ of the branching ratios  for
the considered semileptonic $B_c$ decays, as defined in Eq.~(\ref{eq:rdef1}) and evaluated in this paper. Some previous results
as given  in Refs.~\cite{cpc37-093102,wang09,epjc51-833,z-series,jhep1905-094,mi1,mi2} are also listed as comparison..

From the theoretical  predictions for the branching ratios and the ratios $R_{\eta_c,\jpsi}$  as listed in
Eqs.~(\ref{eq:br11}-\ref{eq:br22}) and Table \ref{tab:br1a} and \ref{tab:br1b},  we find the following points:
\begin{itemize}
\item[(1)]
The theoretical predictions for the branching ratios of all considered $B_c \to (\eta_c,\jpsi ) l^- \bar{\nu}_l $ decays  in both the PQCD and ``PQCD+Lattice" approach
 agree well within errors ( around $30\%$ in magnitude).
 Numerically,  the theoretical predictions for a fixed decay mode will  become a little smaller  by a degree of $(5-16)\%$  when the Lattice QCD results
 for the form factors $(f_{0,+},V, A_1)$ are taken into account in the extrapolation of the relevant form factors to higher $q^2$ region.

\item[(2)]
The theoretical predictions for the ratios $R_{\eta_c}$ and $R_{\jpsi}$  in both the PQCD and ``PQCD+Lattice'' approach agree very well,
and  have very  small errors ( less than $5\%$ in magnitude) due to the strong cancellation between the errors of the theoretical predictions for
branching ratios.
Although the theoretical predictions for  $R_{\jpsi} $  as listed in Table  \ref{tab:br1b}  in both the PQCD and ``PQCD+Lattice'' approaches
are still smaller than the measured value $0.71\pm 0.24$  as reported by LHCb Collaboration \cite{lhcb-18a},  but still agree with it
because of the still large errors of the experimental measurements.
We believe that the ratios $R_{\eta_c}$ and $R_{\jpsi}$ could be measured in high precision at the future LHCb experiment and can help us
to test the theoretical models or approaches.

\item[(3)]
Although  the theoretical predictions for the decay rates from different methods or approaches can be rather different, even reaches a factor of two or three,
the theoretical predictions for the ratios $R_{\eta_c}$ and $R_{\jpsi}$  from different works
~\cite{cpc37-093102,wang09,nrqcd,epjc51-833,z-series,jhep1905-094},  however,  agree very well with each other within $30\%$ of the central value.

\end{itemize}

For both kinds of the semileptonic decays $B \to   D^{(*)} l^- \bar{\nu}_l $ and $B_c^-  \to (\eta_c,\jpsi ) l^- \bar{\nu}_l$,  their  quark level  weak decays
are indeed  the same charged current tree transitions:  $b \to c  l^- \bar{\nu}_l $ with $l=(e,\mu,\tau)$.
The only difference between them is  the spectator quark:  one is the heavy charm quark, another is the light up or down quark.
Consequently,  it is reasonable to assume  that the dynamics for these two kinds of semileptonic decays are similar in nature,
we therefore can use similar method  to study these two kinds of semileptonic decays.

For $B \to   D^{(*)}  \tau \bar{\nu}_\tau $ decays,  besides the decay rates and the ratios $R( D^{(*)})$,   the longitudinal polarization
$P_{\tau}(D^{(*)})$ of the tau lepton  and  the fraction of $D^*$ longitudinal polarization  $F_L^{D^*}$  are also the additional physical observables  and
sensitive to some kinds of new physics \cite{ptau1,ptau2,ptau3,ptau4} .
The first measurement of $P_\tau(D^*)$ and $F_L^{D^*}$ have been reported very recently by Belle Collaboration \cite{prl118-801,prd97-012004,1903tau}:
\beq
P_\tau( D^*) &=& -0.38 \pm 0.51(stat.) ^{+0.21}_{-0.16}(syst.),   \label{eq:ptaustar} \\
F_L(D^*) &=&  0.60 \pm 0.08(stat.) \pm  0.04 (syst.).   \label{eq:flstar}
\eeq
They are compatible  with the SM predictions : $P_\tau( D^*) = -0.497 \pm 0.013$ for $\bar{B} \to D^* \tau^- \bar{\nu}_\tau$ \cite{ptau2,ptau4},
and $F_L(D^*) =  0.441 \pm 0.006 $ \cite{huang18} or  $0.457\pm 0.010$ \cite{bhatt18}.

For   $B_c \to (\eta_c,\jpsi) \tau \bar{\nu}_\tau $ decays, we  consider the relevant longitudinal polarizations $P_\tau(\eta_c)$ and $P_\tau(\jpsi)$,
and define them  in the same way as the one  for $P_\tau(D^{(*)})$ in Refs.~\cite{ptau1,ptau2,ptau3,ptau4}:
\beq
P_\tau(X) = \frac{\Gamma^+(X) - \Gamma^-(X)}{ \Gamma^+(X) + \Gamma^-(X) }, \quad {\rm for} \quad X=(\eta_c,\jpsi), \label{`eq:ptaudef1}
\eeq
where $\Gamma^\pm(X)$ denotes the decay rates of $B_c\to  X \tau \bar{\nu}_\tau $ with a $\tau$ lepton helicity $\pm 1/2$.
Following Ref.~\cite{ptau3},  the explicit expressions of $d\Gamma^\pm/dq^2$  for the considered  semileptonic $B_c$ decays here can be written in the following form:
\beq
\frac{d\Gamma^{+} }{dq^2} (B_c  \to \eta_c \tau \bar{ \nu}_{\tau} ) &=& \frac{ G_F^2 |V_{cb}|^2}{ 192\pi^3 m_{B_c}^3} \;  q^2 \sqrt{\lambda(q^2)}
\left( 1 - \frac{m_\tau^2}{ q^2} \right)^2 \frac{m_\tau^2}{ 2q^2} \left ( H_{V,0}^{s\,2} + 3 H_{V,t}^{s\,2} \right) , \label{eq:dg1a}\\
\frac{d\Gamma^{-}}{dq^2}(B_c \to \eta_c \tau  \bar{\nu}_{\tau} ) &=& {G_F^2 |V_{cb}|^2 \over 192\pi^3 m_{B_c}^3}\; q^2 \sqrt{\lambda(q^2)} \left( 1 - {m_\tau^2 \over q^2} \right)^2
\left( H_{V,0}^{s\,2} \right) , \label{eq:dg1b}\\
\frac{d\Gamma^{+}}{dq^2}(B_c  \to \jpsi \tau  \bar{\nu}_{\tau}) &=& {G_F^2 |V_{cb}|^2 \over 192\pi^3 m_{B_c}^3}\; q^2 \sqrt{\lambda(q^2)}
 \left( 1 - {m_\tau^2 \over q^2} \right)^2 {m_\tau^2 \over 2q^2} \non
&& \cdot \left ( H_{V,+}^2 + H_{V,-}^2 + H_{V,0}^2 +3 H_{V,t}^2 \right ) , \label{eq:dg2a} \\
\frac{d\Gamma^{-}}{dq^2}(B_c  \to \jpsi \tau  \bar{\nu}_{\tau}) &=& {G_F^2 |V_{cb}|^2 \over 192\pi^3 m_{B_c}^3}\; q^2 \sqrt{\lambda(q^2)} \left( 1 - {m_\tau^2 \over q^2} \right)^2
\left ( H_{V,+}^2 + H_{V,-}^2 + H_{V,0}^2 \right) ,  \label{eq:dg2b}
\eeq
with the functions $H_i(q^2)$
\beq
H_{V,0}^s(q^2) & =&  \sqrt{\lambda(q^2) \over q^2} f_+(q^2) ,\label{eq:hi11}\\
H_{V,t}^s(q^2) & =& {m_{B_c}^2-m_{\eta_c}^2 \over \sqrt{q^2}} f_0(q^2) , \label{eq:hi12}\\
H_{V,\pm}(q^2) & =& (m_{B_c}+m_{\jpsi}) A_1(q^2) \mp { \sqrt{\lambda(q^2)}\;  V(q^2) \over m_{B_c}+m_{\jpsi} }  , \label{eq:hi13} \\
H_{V,0}(q^2)   & =& { m_{B_c}+m_{\jpsi} \over 2m_{\jpsi}\sqrt{q^2} } \left[ -(m_{B_c}^2-m_{\jpsi}^2-q^2) A_1(q^2)
+ { \lambda(q^2) \;  A_2(q^2) \over (m_{B_c}+m_{\jpsi})^2 } \right] , \label{eq:hi14} \\
H_{V,t}(q^2)   & =& -\sqrt{ \lambda(q^2) \over q^2 } A_0(q^2) , \label{eq:hi15}
\eeq
where  $0\leq q^2 \leq \left (m_{\rm B_c} -m_X \right )^2 $ and $\lambda(q^2) = \left ( m_{\rm B_c}^2+m_X^2-q^2 \right )^2 - 4 m_{\rm B_c}^2 m_X^2$
with $X=(\eta_c,\jpsi)$,
and the explicit expressions of  the form factors $f_{+,0}(q^2)$, $V(q^2)$ and $A_{0,1,2}(q^2)$  in PQCD approach have been given
in Eqs.~(\ref{eq:f0q2},\ref{eq:Vqq}-\ref{eq:A2qq}).

After making the proper integrations over  $q^2$,  we found the theoretical   predictions for the longitudinal polarization $P_\tau$ for the considered
semileptonic $B_c \to (\eta_c,\jpsi) l^- \bar{\nu}_l $ decays :
\beq
P_{\tau}( \eta_c) = 0.37\pm 0.01,  \quad  P_{\tau}(\jpsi) = -0.55 \pm 0.01,  \label{eq:ptau1a}
\eeq
in the PQCD approach, and
\beq
P_{\tau}(\eta_c) = 0.36\pm 0.01,  \qquad  P_{\tau}(\jpsi) = -0.53\pm 0.01 ,  \label{eq:ptau1b}
\eeq
in the  `` PQCD + Lattice" approach. The dominant errors come from the  uncertainty of $\beta_{B_c}$ and $m_c$.
Following the new measurement of   the  longitudinal polarization $P_{\tau}(D^*)$ for $B \to D^*  \tau \nu_\tau $ at Belle \cite{prl118-801},
we believe that the similar longitudinal polarization $P_{\tau}(\eta_c)$ and $P_{\tau}( \jpsi)$  could be measured in the near future
LHCb experiment when enough amount of $B_c$ decay events are collected.


\section*{5.  Summary } \label{sec:5}

In this paper, we studied  the semileptonic decays $B_c \to (\eta_c,\jpsi) l \bar{\nu}$ by employing the pQCD factorization approach with the usage of some new
inputs: (a) we  used the newly defined DAs of the $B_c$ meson instead of the old delta-function; (b)  we used the new BCL parametrization for the extrapolation
of the form factors from low $q^2$ region to $q^2_{max}$; and (c) we  take into account currently known lattice QCD results of form factors at several points
as the new input in our fitting procedure.
We calculated the form factors $f_{\rm 0,+}(q^{\rm 2})$, $V(q^{\rm 2})$ and $A_{\rm 0,1,2}(q^{\rm 2})$ of  the
$B_c \to (\eta_c,\jpsi)$ transitions,  presented  the  predictions for the  branching ratios ${\cal B}(B_c  \to
 (\eta_c,\jpsi) l  \bar{\nu}_l)$ ,  the ratios  $R_{\eta_c}$ and $R_{\jpsi}$ of the branching ratios,  and the longitudinal polarizations $P_\tau(\eta_c)$ and $P_\tau(\jpsi)$
 of the final $\tau$ lepton.

From the numerical calculations and phenomenological analysis we found the following points:
\begin{enumerate}
\item[(1)]
The theoretical predictions for the branching ratios of  $B_c \to (\eta_c,\jpsi) l \bar{\nu}$ decays from both PQCD and ``PQCD+Lattice" approach agree
very well with each other,   a small decrease by about $(5-16)\%$  will be produced when the lattice QCD input
 for the form factors $(f_{0,+}(8.72),V(5.44), A_1(10.07))$ are taken into account in the extrapolation of the relevant form factors to higher $q^2$ region.

\item[(2)]
The theoretical predictions for the ratios $R_{\eta_c}$ and $R_{\jpsi}$ are the following:
\beq
R_{\rm \eta_c}&=&0.34\pm 0.01, \quad   R_{\rm \jpsi}=0.28\pm 0.01, \quad {\rm in \ \ PQCD}, \\
R_{\rm \eta_c}&=&0.31\pm 0.01 , \quad   R_{\rm \jpsi}=0.27\pm 0.01, \quad {\rm in \ \ PQCD+Lattice},
\eeq
The central values of above theoretical predictions for $R_{\jpsi}$ are smaller than the measured one as shown in Eq.~(\ref{eq:rdata1}),
but still agree with it within the errors.

\item[(3)]
The theoretical predictions for  the longitudinal polarization  $P(\tau)$ of the tau lepton are the following:
\beq
P_{\tau}( \eta_c) &=& 0.37\pm 0.01,  \quad  P_{\tau}(\jpsi) = -0.55 \pm 0.01,   \quad {\rm in \ \ PQCD}, \\
P_{\tau}( \eta_c) &=&  0.36\pm 0.01,  \quad P_{\tau}(\jpsi) = -0.53\pm 0.01 , \quad {\rm in \ \ PQCD+Lattice}.
\eeq
These predictions could be tested in the near future LHCb experiments.

\end{enumerate}


\begin{acknowledgments}

We wish to thank Wen-Fei Wang and Ying-Ying Fan for valuable discussions. This work was supported by the National Natural Science Foundation of
China under Grant  No.~11775117 and 11235005.

\end{acknowledgments}


\appendix

\section{Relevant functions}

In this appendix, we present the explicit expressions for some functions appeared in the previous sections.
The hard functions $h_1$ and $h_2$ appeared in Eq.~(\ref{eq:hiti})  can be written as
\beq
h_1&=&K_0(\beta_1 b_1) \left [ \theta(b_1-b_2)I_0(\alpha_1b_2)K_0(\alpha_1b_1) +\theta(b_2-b_1)I_0(\alpha_1b_1)K_0(\alpha_1b_2) \right ], \non
h_2&=&K_0(\beta_2 b_2) \left [\theta(b_1-b_2)I_0(\alpha_2b_2)K_0(\alpha_2b_1) +\theta(b_2-b_1)I_0(\alpha_2b_1)K_0(\alpha_2b_2) \right ], \ \
\label{eq:h1h2}
\eeq
with
\beq
\alpha_1 &=& m_{B_c}\sqrt{2rx_2\eta+r^2_b-1-r^2x^2_2} , \non
\alpha_2&=&m_{B_c}\sqrt{rx_1\eta^++r^2_c-r^2} , \non
\beta_1  &=& \beta_2 = m_{B_c}\sqrt{x_1x_2r\eta^+-r^2x^2_2},
\eeq
where $r_q=m_q/m_{B_c}$ with $q=(c,b)$,  $r=m_{\eta_c}/m_{B_c}$ ( $r=m_{\jpsi}/m_{B_c}$) when it appears in the form factors $f_{+,0}(q^2)$ (  $V(q^2)$ and
$A_{0,1,2}(q^2)$).  $\eta$ and $\eta^+$ are defined in Eq.~(\ref{eq:eta}).
 The functions $K_0$ and $I_0$ in Eq.~(\ref{eq:h1h2}) are  the modified Bessel functions.
The term inside the square-root symbol  of  $\alpha_{(1,2)}$ and $\beta_{(1,2)}$
may be positive or negative.  When such term is negative,  the argument of  the functions $K_0$ and $I_0$  becomes imaginary,
and the associated Bessel functions  $K_0$ and $I_0$ will consequently transform in the following way
\beq
K_0(\sqrt{y})|_{y<0} &=& K_0(i\sqrt{|y|})= \frac{i \pi}{2} [J_0(\sqrt{|y|}) + i Y_0(\sqrt{|y|})] \non
I_0(\sqrt{y})|_{y<0}  &=& J_0(\sqrt{|y|}) ,
\eeq
where the functions $J_0(x)$ and $Y_0(x)$ can be written in the following form\cite{sf8e}
\footnote{One can find the expression of $J_0(x)$ and $Y_0(x)$ in Sec.8.411 and 8.415 of Ref.~\cite{sf8e}.}
\beq
J_0(x)&=& \frac{1}{\pi}\int_{0}^{\pi} \cos(x\sin{\theta})\; d\theta, \quad (x >0) \non
Y_0(x)&=& \frac{4}{\pi^2}\int_{0}^{1} \frac{\arcsin(t) }{\sqrt{1-t^2} }\sin(xt) dt -
\frac{4}{\pi^2}\int_{1}^{\infty } \frac{\ln \left ( t+\sqrt{t^2-1} \right) }{\sqrt{t^2-1} } \sin(xt) dt , \quad  (x>0).
\eeq

The factor $\exp[-S_{ab}(t)]$ in Eq.~(\ref{eq:hiti}) contains the Sudakov logarithmic corrections and
the renormalization group evolution effects of both the wave functions and the
hard scattering amplitude with $S_{ab}(t)=S_{B_c}(t)+S_X(t)$  as given in Ref.~\cite{Bc-am}
\beq
S_{B_c}&=&s_c\left(\frac{x_1}{\sqrt{2}}m_{B_c}, b_1\right)+\frac{5}{3}\int^t_{m_c}\frac{d\bar\mu}{\bar\mu} \gamma_q(\alpha_s(\bar\mu)), \non
S_{\eta_c}&=&  s_c\left(\frac{x_2}{\sqrt{2}}m_{\eta_c}\; \eta^+,b_2\right) + s_c\left( \frac{(1-x_2)}{\sqrt{2}} m_{\eta_c}\; \eta^+, b_2\right)+
2\int^t_{m_c}\frac{d\bar\mu}{\bar\mu} \gamma_q(\alpha_s(\bar\mu)), \non
S_{J/\psi}&=&s_c\left( \frac{x_2}{\sqrt{2}}m_{\jpsi}\; \eta^+,b_2\right)
+s_c\left(\frac{(1-x_2)}{\sqrt{2}} m_{\jpsi}\; \eta^+,  b_2\right)+ 2\int^t_{m_c}\frac{d\bar\mu}{\bar\mu} \gamma_q(\alpha_s(\bar\mu)),
\eeq
where $\eta^+$ is defined in Eq.~(\ref{eq:eta}), while  the hard scale $t$ and the quark anomalous dimension $\gamma_q=-\alpha_s/\pi$,
which  governs the aforementioned renormalization group evolution.
The Sudakov exponent $s_c(Q,b)$ for an energetic charm quark is expressed \cite{Bc-am} as the difference
\beq
s_c(Q,b)&=&s(Q,b)-s(m_c,b) \non
&=&\int_{m_c}^Q\frac{d \mu}{\mu}
\left[\int_{1/b}^{\mu}\frac{d\bar\mu}{\bar\mu}A(\alpha_s(\bar\mu))
+B(\alpha_s(\mu))\right].
\eeq
The hard scales $t_i$ are chosen as the
largest scale of the virtuality of the internal particles in the hard $b$-quark decay diagram,
\beq
t_1&=&\max\{\alpha_1, 1/b_1, 1/b_2\},\non
t_2&=&\max\{\alpha_2, 1/b_1, 1/b_2\}.
\eeq


\end{document}